\documentclass[fleqn,10pt]{wlscirep}
\usepackage[utf8]{inputenc}
\usepackage[T1]{fontenc}
\usepackage{xspace}
\usepackage{xcolor}
\usepackage{scalerel}

\newcommand{\Jzero}{SDSS\ J0919$+$2720\xspace}

\newcommand*\arcsec{\ensuremath{^{\prime\prime}}}
\newcommand*\arcmin{\ensuremath{^{\prime}}}
\newcommand*\farcs{\ensuremath{\overset{\prime\prime}{.}}}
\newcommand{\lenstro}{{\tt Lenstronomy}}
\newcommand{\slit}{{\tt SLITronomy}}
\newcommand{\angstrom}{\mbox{\normalfont\AA \ }}
\def\ksmpc{${\, \mathrm{km}\, \mathrm{s}^{-1}\, \mathrm{Mpc}^{-1}}$\xspace}
\def\ks{${\, \mathrm{km}\, \mathrm{s}^{-1}}$\xspace}
\newcommand{\lcdm}{$\mathrm{\Lambda CDM}$\xspace}

\title{Strong gravitational lensing by AGNs as a probe of
the quasar-host relations in the distant Universe}

\author[1,2,*]{Martin Millon}
\author[1]{Fr\'ed\'eric Courbin}
\author[1,3]{Aymeric Galan}
\author[4]{Dominique Sluse}
\author[5]{Xuheng Ding}
\author[6]{Malte Tewes}
\author[7]{S. G. Djorgovski}
\affil[1]{Institute of Physics, Laboratory of Astrophysics, Ecole Polytechnique 
F\'ed\'erale de Lausanne (EPFL), Observatoire de Sauverny, 1290 Versoix, 
Switzerland}
\affil[2]{Kavli Institute for Particle Astrophysics and Cosmology and Department of Physics, Stanford University, Stanford, CA 94305, USA}
\affil[3]{Technical University of Munich, TUM School of Natural Sciences, Department of Physics, James-Franck-Straße 1, 85748 Garching, Germany}
\affil[4]{STAR Institute, Quartier Agora, All\'ee du six Aout 19c, 4000 Li\`ege, Belgium}
\affil[5]{Kavli Institute for the Physics and Mathematics of the Universe, The University of Tokyo, Kashiwa, Japan 277-8583 (Kavli IPMU, WPI)}
\affil[6]{Argelander-Institut f\"ur Astronomie, Universit\"at Bonn, Auf dem H\"ugel 71, 53121 Bonn, Germany}
\affil[7]{California Institute of Technology, Pasadena, CA 91125, USA.}

\affil[*]{corresponding author(s): Martin Millon (martin.millon@stanford.edu)}

\begin{abstract}
The tight correlations found between masses of supermassive black holes (SMBHs) and the luminosities, total stellar masses, and velocity dispersions of their host galaxies are often interpreted as a sign of their co-evolution. Studying these correlations across redshift provides a powerful insight into the evolutionary path followed by the quasar and its host galaxy. While the mass of the black hole is accessible from single-epoch spectra, measuring the mass of its host galaxy is challenging as the active nucleus largely overshines its host. Here, we present a technique to probe quasar-host relations beyond the local universe with strong gravitational lensing, hence overcoming the use of stellar population models or velocity dispersion measurements, both prone to degeneracies. We study in detail one of the three known cases of strong lensing by a quasar to accurately measure the mass of its host and to infer a total lensing mass of $\log_{10}(M_{\rm Tot, h}/M_{\odot}) = 10.27^{+0.06}_{-0.07}~$ within the Einstein radius of 1.2 kpc. The lensing measurement is more precise than any other alternative technique and compatible with the local $M_{BH}$-$M_{\star, h}$ scaling relation. The sample of such quasar-galaxy or quasar-quasar lensing systems should reach a few hundreds with Euclid and Rubin-LSST, thus enabling the application of such a method with statistically significant sample sizes. 

\end{abstract}
\begin{document}

\flushbottom
\maketitle

\thispagestyle{empty}

\section*{Introduction}
\begin{figure}[h!]
    \centering
    \includegraphics[width=\textwidth]{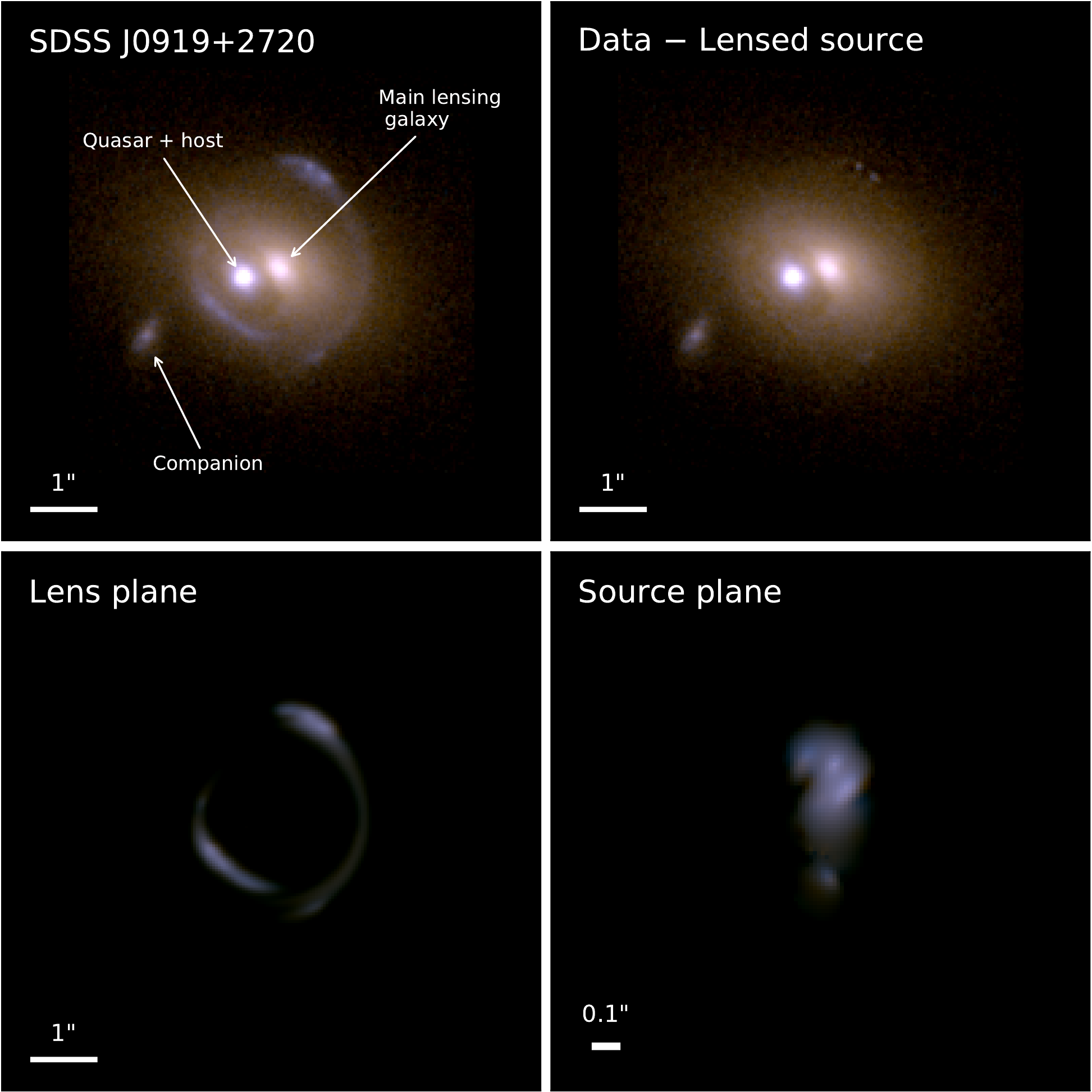}
    \caption{The \Jzero strongly lensed system. \textit{Upper left: } Color composite of the HST images in the F475W and F814W bands. The different components of the system are indicated: the main lensing elliptical galaxy at $z = 0.209$ produces most of the lensing effect on the (blue) star-forming galaxy at $z = 0.558$. The bright blue quasar and its host galaxy, also at $z = 0.209$, act as a secondary deflector for which we can measure the total mass. \textit{Upper right: } HST image, where the best-fit model of the gravitational arcs has been subtracted.
    \textit{Lower left:} Best-fit model of the lensed arcs.
    \textit{Lower right:} Source reconstruction using wavelet decomposition and sparse regularization. The pixel size is $0\farcs013$, i.e. three times smaller than the drizzled HST images (see Method section for more detail on the lens modeling).}
    \label{fig:rgb}
\end{figure}

We present an exceptional case of strong lensing by an elliptical galaxy displaying a well-visible substructure pointed to us by a quasar: \Jzero \citep[][]{Courbin2012}. The elliptical galaxy and \Jzero share the same redshift, $z_{l} = 0.209$, are separated by less than 1\arcsec\ on the plane of the sky and act as a gravitational lens on a distant star-forming source at $z_{s} = 0.558$. Most of the lensing effect is produced by the main elliptical lens galaxy (see Fig.~\ref{fig:rgb}), but the quasar and its host galaxy, even hidden in the glare of the quasar light, produce detectable lensing signal. We take advantage of this fortunate lensing event and of the fact that the quasar and the main elliptical lensing galaxy do not share the same location, to measure the total (dark + luminous) mass of a quasar host galaxy with strong gravitational lensing.

Quasars are known to follow tight correlations between the mass of their central supermassive black hole (SMBH) and the fundamental properties of their host galaxy, such as the stellar mass $M_{\rm \star, h}$ and stellar velocity dispersion $\sigma_{\rm \star, h}$ related to the total mass $M_{\rm Tot, h}$. The very existence of these scaling relations suggests a connection between the quasar activity and the formation of its host galaxy \citep[e.g.][]{Ferrarese2000, Gebhardt2000}. The physical process leading to these correlations remains unclear since the gravitational influence of the SMBH is limited to parsec scales while the typical scale of their host galaxy is three orders of magnitude larger. On the one hand, AGN (Active Galactic Nuclei) feedback in numerical simulations seems to reproduce the observed correlations \citep[e.g.][]{Springel2005, Dimatteo2008} but, on the other hand, they could also simply result from the hierarchical assembly of multiple mergers \citep[e.g.][]{Peng2007}. Discriminating between these scenarios requires measuring the bulge mass of quasar host galaxies not only in the local universe but also at higher redshift where velocity dispersion measurements are challenging but where lensing is both precise and accurate. If the scatter in the $M_{\rm BH} - M_{\rm Tot, h}$ relation increases with redshift, this would support the hierarchical assembly scenario \citep{Kormendy2013}.  

Moreover, it is not well established if there is an offset between the relations observed in the local and high-redshift Universe. For example, Sexton et al. (2019) \cite{Sexton2019} did not find any evidence for an evolution of the $M_{\rm BH}-\sigma_{\rm \star, h}$ relation with redshift. On the other hand, recent works by Ding et al. (2020, 2021)\citep{Ding2020, Ding2021} found a positive evolution of the $M_{\rm BH}/M_{\star, h}$ ratio, compatible with a scenario where the SMBH would grow at an earlier time and the bulge of its host galaxy is catching up later. If both of these results are correct, this would imply that the bulge stellar mass increases without significantly changing the total mass of the galaxy. Possible mechanisms leading to the growth of the bulge without increasing the black hole mass nor the total mass would involve a transfer of stellar mass from the disk to the bulge through minor mergers or disk instabilities \citep{Jahnke2009, Schramm2013}. The absence of evolution in the $M_{\rm BH}-\sigma_{\rm \star, h}$ relation \citep[e.g.][]{Sexton2019, Schulze2014} suggests that the offset found in earlier works \cite[e.g.][]{Woo2008, Treu2007} could result from selection effects on both $M_{\rm BH}$ and $\sigma_{\star, h}$. In that case, the absence of change in these scaling relations across the cosmic time would thus indicate a close co-evolution between the SMBH and its host, possibly regulated through AGN feedback or because they share a common gas reservoir \citep{Cen2015}. 

All these studies are currently limited by the difficulty to measure $M_{\star, h}$ and $\sigma_{\rm \star, h}$ at high redshift because of deblending issues between the quasar and stellar emission, uncertainties on the initial mass function (IMF), and selection effects due to the luminosity-selection process involved in most quasar samples \citep[e.g][]{Lauer2007}. Additionally, converting $\sigma_{\rm \star, h}$ into $M_{\rm Tot, h}$ requires breaking the mass-anisotropy degeneracy. This can be achieved with spatially resolved kinematics measurements but these are very difficult in the distant universe because the quasar is often brighter than the stellar component. In summary, measuring the total and stellar mass of the host both faces important observational challenges and relies on questionable assumptions. In this paper, we propose a technique based on gravitational lensing to measure the total mass of the host with excellent precision, which also imposes a strict upper limit on the stellar mass within the Einstein radius.

\Jzero consists of a test bench to carry out the experiment, at redshift $z = 0.209$. It is one of the three known cases of lensing by a quasar discovered by Courbin et al. (2012) \cite{Courbin2012}. Other candidates have been proposed in early searches by Claeskens et al. (2000) \citep{Claeskens2000} and more recently by Meyer et al. (2019) \citep{Meyer2019} but they have not yet been confirmed with high-resolution imaging. Among these three confirmed systems, \Jzero is the only Type-I AGN displaying well-visible gravitational arcs. This paper presents the analysis of such a system, where the lensing (total) mass, the host stellar mass, and the black hole mass are measured. \Jzero is, for now, the most striking case of lensing by a quasar, but a few hundreds will be found in the near future with wide-field imaging surveys \citep[][]{Taak2020}, opening the path to measure robust scaling relations up to redshift close to $z=1$, i.e. well beyond the local universe. Although \Jzero requires complex lens models to disentangle the mass of the quasar host from the mass of the nearby massive elliptical galaxy, most future cases of lensing by a quasar should be massive enough to produce an Einstein ring by themselves \citep{Taak2020}, allowing us to obtain a straightforward and model-independent measurement of the lensing mass directly from the Einstein radius. 

Throughout this paper, we assume a flat-\lcdm cosmology with $H_0 = 70$ \ksmpc, $\Omega_m = 0.3$ and $\Omega_{\Lambda}=0.7$ to compute the angular diameter distances from the redshifts measurements, and we adopt the AB magnitude system.

\section*{Results}
\subsection*{Lensing mass estimates}
We use HST images (program GO-12233; PI Courbin) of \Jzero taken in the F475W and F814W optical bands to reconstruct the total mass distribution of the system. Because the positions of the quasar host and that of the main lens do not coincide, our lens models are able to separate the total mass of the quasar and its host from that of the main elliptical lensing galaxy. To perform the modeling, we fit simultaneously the two HST bands, imposing that the light and mass profiles in the lens plane share the same center in both bands and fixing their position during the fit.

Following common practice for strong lens models, we describe the mass of the main lens galaxy as a Power-law Elliptical Mass Distribution (PEMD) whose slope, $\gamma_{\rm l}$, can vary during the fit. For the mass of the quasar host galaxy, we consider a Singular Isothermal Ellipsoid (SIE) which by definition has a fixed slope of $\gamma_{\rm host}=2$. We do not consider any explicit point-mass component for the quasar mass as it is negligible in front of the mass of its host. The light of the source galaxy is reconstructed with shapelets, leading to 103 linear parameters, directly optimized from the inversion of the lens equation, and 18 non-linear free parameters simultaneously optimized in a Bayesian framework. Our best-fit model is presented in Fig.~\ref{fig:rgb}. As a final step, we perform a fully non-parametric reconstruction of the source, for the best-fit mass model, on a pixelated grid regularized with wavelets \citep{galan2021}. This leads to a high-resolution image of the star-forming source galaxy. 

As the quasar host is not centered on the main lens galaxy, our models have high sensitivity to its mass and we can measure it in a very robust way. Our best model provides its Einstein radius $\theta_{\rm E, h} = 0\farcs355^{+0.024}_{-0.028}$, which translates into a total mass of $\log_{10}(M_{\rm Tot, h}/M_{\odot}) = 10.27^{+0.06}_{-0.07} ~$ within the Einstein radius of 1.2 kpc in the lens plane. Models generated with and without including the quasar host galaxy have a very different Bayesian Information Criterion (BIC), with $\Delta BIC = 234$ (see Table \ref{tab:BICvalue} and the Methods section for alternative modeling assumptions). This strongly favours models with explicit modeling of the substructure marked by the quasar and shows that we have high sensitivity to it. We also measure the Einstein radius of the main galaxy, $\theta_{\rm E, m} = 1\farcs016^{+0.016}_{-0.016}$, corresponding to a total mass of $\log_{10}(M_{\rm Tot, m}/M_{\odot}) = 11.18^{+0.01}_{-0.01} ~$ within a radius of 3.5 kpc. Our lens model predicts a luminosity-weighted line-of-sight velocity dispersion (as described in \cite{Suyu2010}) of $\sigma_{\star ,h} = 111 \pm 3$ \ks for the quasar host galaxy and of $\sigma_{\star ,m} = 227 \pm 3$ \ks for the main galaxy. These estimates are obtained by solving the spherical Jeans equations, assuming isotropic motion of the stars. They are given here for reference, only as a dynamical equivalent of the lensing measurement.

\begin{figure}[h!]
    \centering
    \includegraphics[width=0.8\textwidth]{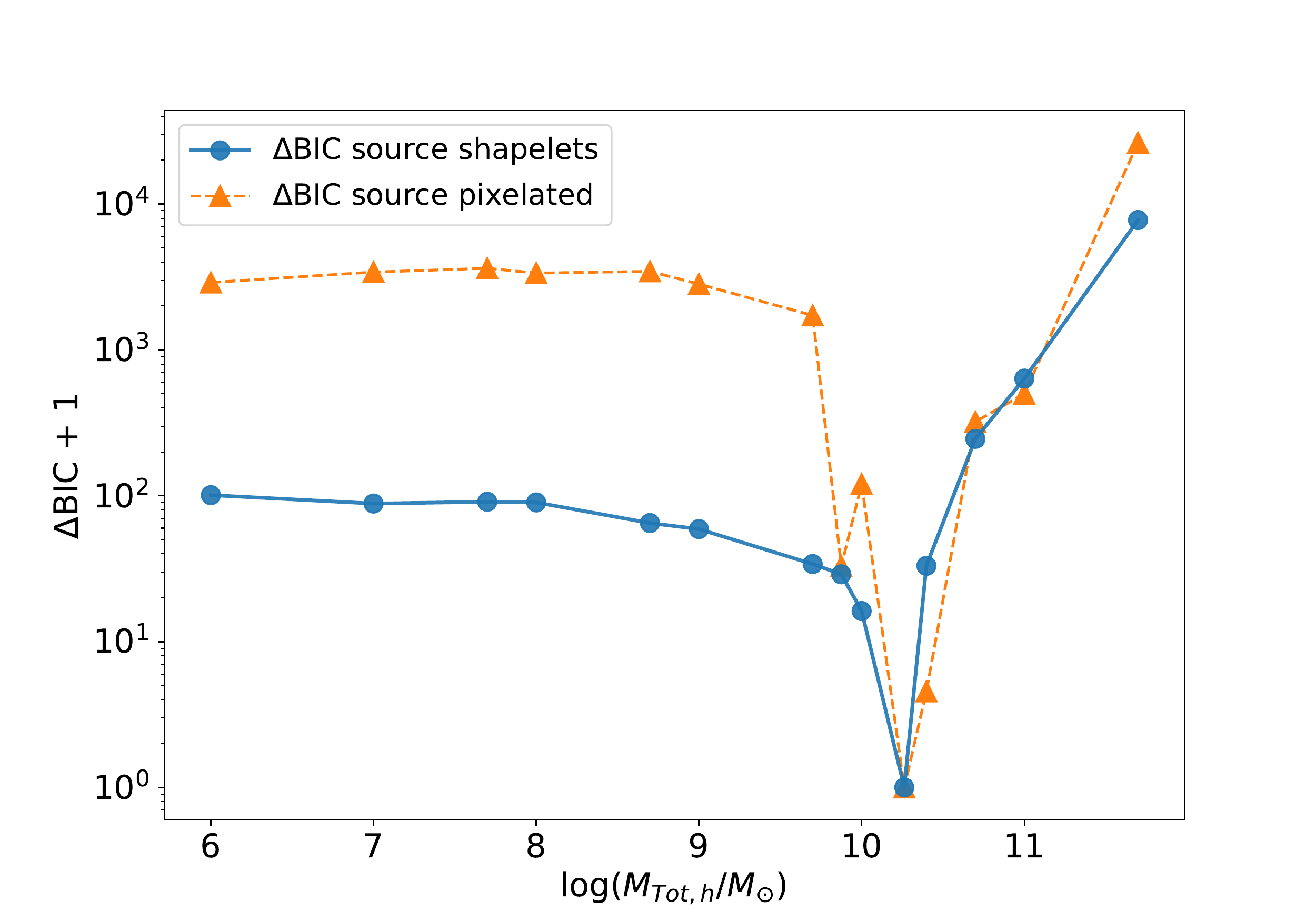}
    \caption{Value of $\Delta$BIC for different masses of the quasar host galaxy, with a clear minimum for host total mass of $\log_{10}(M_{\rm Tot, h}/M_{\odot})=10.27 ~$. We show the result for models with (analytic) shapelets decomposition of the source and fully pixelized source regularized with wavelets. The $\Delta$BIC values are computed relative to the best model for each source reconstruction method. Note that the two corresponding curves do not overlap other than at the minimum because the two types of models have very different numbers of degrees of freedom. Still, the positions of the minima agree well, showing that our results are robust against the source reconstruction method.}
    \label{fig:BIC}
\end{figure}

As an additional test, we verify the robustness of the fit by optimizing a series of models with fixed $M_{\rm Tot, h}$ in the range $10^6 M_\odot < M_{\rm Tot, h} < 10^{12} M_\odot$. For each value of $M_{\rm Tot, h}$ we re-optimize all other free parameters, including the source reconstruction, and we check whether the new model optimization and source reconstruction compensate for the variations introduced in the image plane. We then compute the corresponding value of $\Delta BIC$ to determine whether the changes seen in the image residuals are statistically significant. Figure~\ref{fig:BIC} illustrates this process and shows a minimum in $\Delta BIC$ for $\log_{10}(M_{\rm Tot, h}/M_{\odot})=10.27 $, which also happens to be the value we find when optimizing models where $M_{\rm Tot, h}$ is free. The plateau seen until $M_{\rm Tot, h} < 10^8 M_{\odot}$ can be interpreted as our detection threshold. This plateau provides similar information as the sensitivity maps of Despali et al. (2021) \citep{Despali2021} and indicates that we are sensitive to masses of the order of $10^8-10^9 M_{\odot}$, at the position of the quasar. This is excellent given that the quasar is not located right on the lensed image of the source, where mass sensitivity would be even better.

\subsection*{Stellar and black hole mass estimates}
Strong lensing provides us with $M_{\rm Tot, h}$ but the HST images also give access to the stellar light of the quasar host, which we use to estimate its stellar mass, $M_{\rm \star, h}$. This requires to model and remove the quasar light and therefore to model the instrumental Point Spread Function (PSF) accurately. We do this by using HST images of the open cluster NGC~136, taken close in time to the observations of \Jzero and with the same dithering pattern. This young cluster allows us to build a model of the PSF using blue stars with colors matching that of the quasar and close to its position on the detector. 
 

We fit our PSF model at the quasar position as well as a model for the light component of the main lensing galaxy and of the quasar host galaxy. We measure the quasar host flux in an aperture of $0\farcs71$ in diameter, corresponding to twice the Einstein radius of the quasar host. The modeling of the color of the host (F475W-F814W=1.09 mag) translates into a stellar mass of $\log_{10}(M_{\rm \star, h}/M_{\odot}) = 9.80~ $, assuming a Chabrier IMF and solar metalicity. The typical uncertainty on the stellar mass is 0.2 dex, similar to the work of \cite[][]{Ding2020}. Comparing this number with the total lensing mass within an aperture of $0\farcs71$ (2.4 kpc in diameter in the lens plane), we infer a stellar-to-total mass ratio $M_{\rm \star, h}/M_{\rm Tot, h}$ of $0.34\pm 0.27$. The stellar mass of main galaxy $M_{\rm \star, m}$ is estimated in a similar way, i.e. from the photometry within an aperture of $2\farcs03$, corresponding to twice the Einstein radius of the main lens. We infer $\log_{10}(M_{\rm \star, m}/M_{\odot}) = 10.72~$ for Chabrier IMF.

Finally, we compare the stellar and lensing masses with the black hole mass $M_{\rm BH}$. \Jzero has been observed as part of the Sloan Digital Sky Survey (SDSS) and with the Low Resolution Imaging Spectrometer (LRIS) on the Keck telescope \citep{Courbin2012}. Both spectra show prominent broad emission lines in the optical, making it possible to estimate the virial mass of the central black hole via the so-called "single epoch method". Using the measured width of the broad H$\alpha$ line and the continuum flux at 6200\AA, we found a black hole mass of $\log_{10}(M_{\rm BH}/M_{\odot}) = 7.32 ~$ from the Keck spectrum and $\log_{10}(M_{\rm BH}/M_{\odot}) = 7.29 ~$ from the SDSS spectrum, with 0.35 dex uncertainties.

\begin{figure}[h!]
    \centering
    \includegraphics[width=0.75\textwidth]{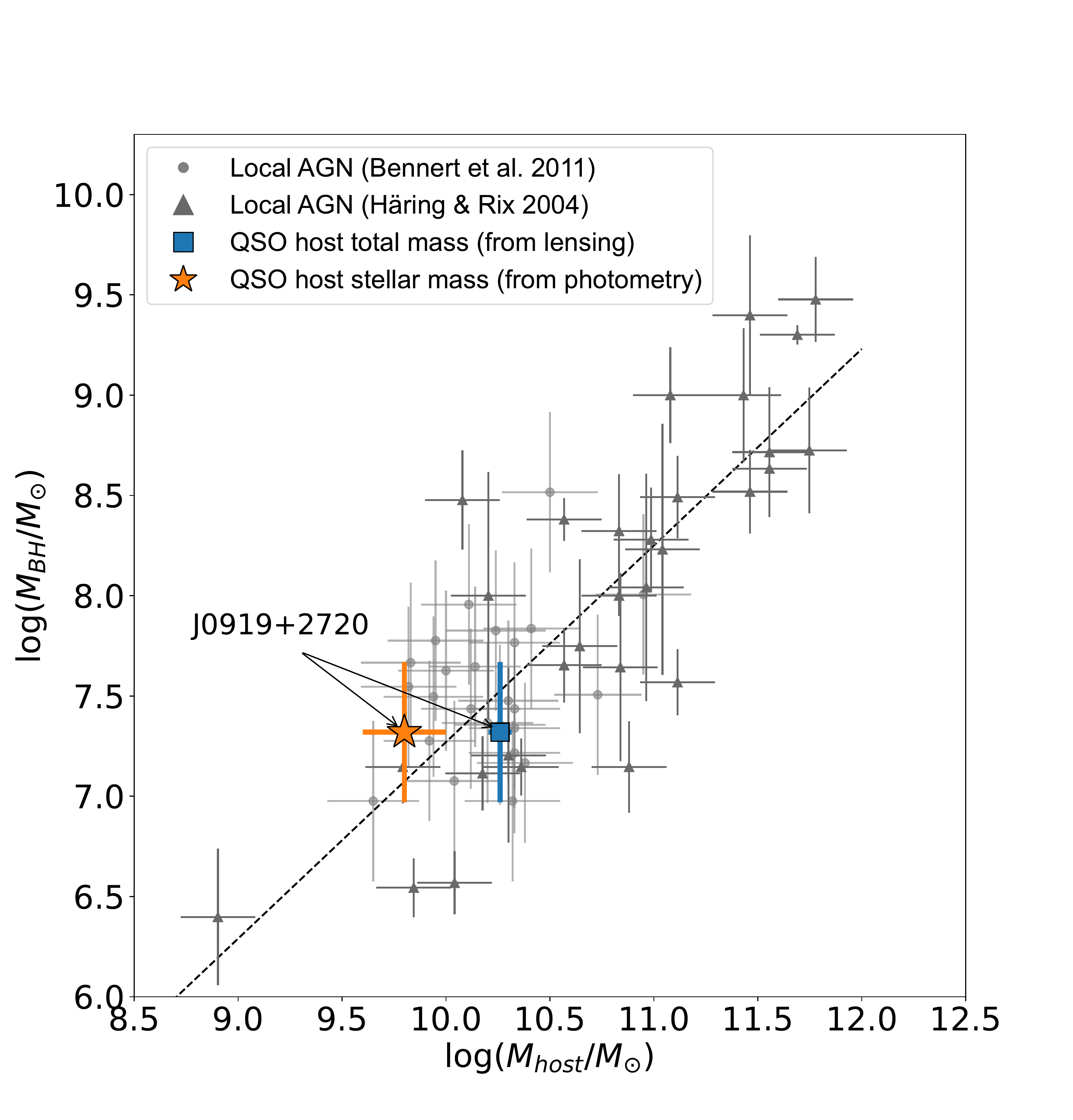}
    \caption{$M_{\rm BH}-M_{\rm \star, h}$ relation for local ($z \lesssim 0.08$) AGNs taken from H\"aring \& Rix (2003)\cite{Haring2004} and Bennert et al. 2011 \cite{Bennert2011}. We show our total mass estimates from lensing and our stellar mass estimates from photometry (assuming Chabrier IMF) for the quasar host galaxy in the \Jzero system. Note that the error bars for the lensing mass estimates (blue symbols) are too small to be displayed.}
    \label{fig:Mbhrelation}
\end{figure}

Figure~\ref{fig:Mbhrelation} shows how \Jzero compares to other local AGNs in the $M_{\rm BH}-M_{\rm \star, h}$ plane. The points corresponding to the stellar mass inferred from photometry of the quasar host or from lensing are well on the correlation. Both masses are measured in an aperture of 2.4 kpc, matching the Einstein radius of the quasar host. The stellar mass and total mass estimates do not differ by more than a factor of 3. 

\section*{Discussion}

Scaling relations between the mass of quasars and that of their host galaxies are a powerful diagnostic of their co-evolution. A change of the relations with redshift may well indicate a change in time of the physical processes at play in quasar and galaxy formation and co-evolution \citep{Kormendy2013}. But for this to work, scaling relations must be established reliably, ideally with one single method and over a broad redshift range.

Building scaling laws is notoriously difficult as this involves the measurements of the mass of the central black hole from emission line properties of the quasar and the mass of the host galaxy. The black hole mass estimate involves assumptions on the geometry and dynamics of the emitting material producing the lines and, as stated in \citep{Kormendy2013} "measuring the mass of the host galaxy is trickier than it sounds". Using multi-band photometry allows us to estimate its stellar mass with fairly large error bars as this requires assumptions on the stellar populations of the host, its metallicity, and a choice of the IMF. Turning the stellar mass into a total mass further assumes a mass-to-light ratio, not to mention that this ratio may well be spatially variable across the galaxy. 

Alternatively, the total mass can be estimated through a stellar velocity dispersion measurement, provided the host galaxy can be seen in the glare of the much brighter quasar. Although this has been attempted in the past with HST and ground-based Adaptive Optics \citep{Grier2013} or even in spectroscopy with VLT \citep{Letawe2007}, the method becomes very difficult with increasing redshift. In addition, the interpretation of the velocity dispersion itself depends on all the above assumptions on photometric stellar masses but it also requires assumptions on the light profile of the host galaxy. For this reason, scaling relations using velocity information have been limited so far to local or intermediate redshift (z $\sim$ 0.5) quasar host galaxies \citep[see e.g.][]{Sexton2019}. 

Strong gravitational lensing offers an extremely valuable alternative to measure the total mass of quasar host galaxies independent of any of the above assumptions: lensing measures the total mass in the Einstein radius, i.e. a very well-defined aperture. \Jzero consists in the first clear case of a quasar acting as a gravitational lens on a more distant galaxy. We test a range of lensing models and find that our mass estimates are robust and little affected by systematic errors. We measure a total mass of $\log_{10}(M_{\rm Tot, h}/M_{\odot}) = 10.27^{+0.06}_{-0.07} ~$ within 2.4 kpc with 16\% precision. An additional source of error may come from line-of-sight (LOS) structures. We could potentially misattribute the mass within the Einstein radius to the host galaxy whereas it comes from a LOS structure. However, it has been demonstrated on a sample of time-delay lenses by the TDCOSMO collaboration that this effect does not exceed a few percent and can be accurately corrected \citep{Millon2020}. Moreover, the lensed sources in this sample are at a much higher redshift ($z_s$ = 1.5-2.5) than our source galaxy at redshift 0.55. These time-delay systems are likely much more affected by LOS effects than our low redshift lensed source. In the case of \Jzero, we can expect that LOS effects are negligible and gravitational lensing is thus much more precise than any other existing techniques. Our measurement is well compatible with scaling relations established in the local universe and, given the small size of the Einstein radius, we measure mostly the bulge total mass. We note that lensing is probing a different quantity than $M_{\star, h}$ or even $\sigma_{\star, h}$ since the lensing mass is sensitive to the \textit{total} projected mass in a cylinder. It can be related to $M_{\star, h}$ by assuming a mass-to-light ratio or to $\sigma_{\star, h}$ by solving the Jeans equation, assuming spherical symmetry of the mass distribution. Consequently, comparing how this lensing mass scales with the black hole mass is well complementary to the other scaling relations and will help to build a consistent picture among the different galaxy evolution scenarios.

As lensing is not affected by the same errors as traditional techniques, it might solve the question of the presence or not of an evolution of the $M_{BH} /M_{Tot,h}$ ratio over redshifts. Such evolution was not detected in the $M_{BH} - \sigma_{\star}$ by \citep{Sexton2019}, up to redshift $\sim 0.6$, contradicting the results found by previous studies \citep{Woo2008, Treu2007}. Lensing could settle this debate by measuring the total mass of the host on a similar redshift range, hence discriminating between 1- a scenario where the black hole grows first and is caught up by its host galaxy to end up on the relation observed locally, and 2- a scenario where the two components co-evolve together, with tight control of the growth of the host, possibly via AGN feedback. We, therefore, propose to use strong lensing to establish the scaling laws involving the total mass of the host galaxy. Not only this will complement other techniques based on pure stellar mass measurements but it will enable us to trace the evolution of the stellar-to-total mass ratio of quasar host galaxies. 

\Jzero is only one case of strong lensing by a quasar, but the present detailed study with sharp HST images lends considerable hope to build scaling relations using reliable total masses with no strong astrophysical assumptions and at redshift typical for lensing galaxies, i.e. up to $z=1$. We emphasize that this method does not require the special geometrical configuration of \Jzero, where the quasar is seen as a lensing substructure of the main deflector. This configuration simply enlarges the Einstein radius and increases the lensing magnification of the background source, which has made this system easier to detect. However, more massive quasar host galaxies are able to produce a detectable Einstein ring by themselves. Recent work \cite{Taak2020} predicts the discovery of $~80$ such systems in the HSC/wide survey and several hundreds of them in future large sky surveys such as Euclid and Rubin-LSST. Euclid will certainly be the best place to look for such objects are the instrumental PSF of 0$\farcs$2 is 2 to 3 times smaller than the typical Einstein radius of a (quasar host) galaxy, hence drastically improving the light contrast between the foreground quasar and the background lensed source. Euclid will also have a sufficient image resolution to obtain precise mass estimates directly from the survey data. The follow-up data needed to study $M_{\rm BH}-M_{\rm Tot, h}$ relation with this technique will therefore be reduced to a single spectrum to measure the black hole mass, which can easily be obtained from ground-based 3m class telescopes. 

When a sample of QSO lenses becomes available, the main limitation in studying the $M_{\rm BH}-M_{\rm Tot, h}$ relation will be the uncertainty on the black hole mass, which currently reaches $\sim$ 0.3 dex. The uncertainty on the total mass obtained with lensing will be negligible in comparison. However, the accuracy of black hole mass measurements in the distant Universe will continue to improve as more measurements become available from reverberation mapping \cite[e.g.][]{Park2017, Bahk2019}. These estimates can then be used to calibrate measurements from single-epoch spectroscopy. Finally, we expect that a large sample of quasar-galaxy and quasar-quasar lenses, comparable in size to all previous samples with velocity dispersions, will be accessible in the next decade and will allow us to robustly determine the $M_{\rm BH}-M_{\rm Tot, h}$ relation across cosmic time.

\newpage
\section*{Methods}
\subsection*{HST observations, data reduction and PSF measurement}
Our HST data consists of six consecutive dithered exposures in the F475W and F814W filters with the WFC3 instrument on board HST. The total exposure time of the combined drizzled frames is 2274s and 2382s in the F475W and F814W bands respectively, with a pixel scale of $0\farcs04$. The noise level at each pixel position is estimated from a Poisson noise component, scaled by the exposure map, and a background noise component, estimated from an empty region of the image with {\tt Sextractor}\citep{Bertin1996}. The PSF is not built from the stars in the field of view of \Jzero, but rather from the observation of the open star cluster NGC136 that has been observed for one full orbit 4 months before \Jzero. Blue stars with the same color as \Jzero fall close to the quasar position on the CCD camera and allow us to reconstruct an exquisite PSF with properties very close to the PSF of the quasar. This is crucial to deblend the quasar host galaxy from the point-source component. Three long exposures of 350s each and one short exposure of 60s were taken in both filters. We use {\tt Sextractor} on the dithered images to perform the photometry of all sources and we select the stars within 1\arcmin\ from the position on the detector of \Jzero. We then keep all stars that have an F475W - F814W color matching that of the quasar within 0.5 mag. We further restrict our selection to sources in the magnitude range 17.5$-$24.5 (18$-$25) in the F814W (F475W) band. We ensure that the selected sources are point-like objects by removing all sources with ellipticity larger than $\epsilon = 0.9$ and a full width at half maximum (FWHM) larger than 2.8 pixels ($0\farcs11$). Finally, we visually inspect the cutouts of the selected stars and remove any star with companions. This results in respectively 12 and 9 useful stars in the F475W and F814W bands. The stars are then stacked using the publicly available Python package \href{https://github.com/sibirrer/AstroObjectAnalyser}{\tt AstroObjectAnalyser}( \url{https://github.com/sibirrer/AstroObjectAnalyser}) to obtain a high signal-to-noise model of the PSF as well as an estimate of the PSF errors at each pixel position. 

\subsection*{Lens modeling}
We use the Python strong lens modeling public package \href{https://github.com/sibirrer/lenstronomy}{\lenstro} \cite{lenstronomy2021} that implements parametric models for the mass distribution of the lens as well as flexible shapelet-based reconstruction of the source. We fit the images in the F475W and F814W bands jointly, i.e. the mass profiles in the models are the same in both bands.

We model the lens light with two analytical Sérsic models at the position of the main lens galaxy sharing the same centroid position. The quasar light emission is modeled with a point source plus 2 Sérsic profiles at the position of the quasar host. The centroid of the two Sérsics are tied together and to the point source position.  We add a Sérsic profile to fit the light of the companion galaxy lying at the South-East of the lens. As a first step, we do not attempt to model the gravitational arcs and fit only the light of the lensing objects. This is used as a starting point for all other lens models. We note that this parametric reconstruction of the lens light is not completely sufficient to model accurately the inner part of the system. \Jzero has a very complex luminous structure, probably due to absorption by dust. However, this occurs only in the central part of the lens galaxy, so we place a mask of $0\farcs84$ in diameter at the center of the frame. 

The gravitational arcs are modeled using an elliptical power-law (PEMD) profile at the position of the main galaxy, along with external shear. The main lens galaxy is a massive early-type galaxy similar in term of morphology, lensing mass, and redshift to the lens galaxies of the SLACS survey \citep{Bolton2006}. We expect that the main deflector also shares a comparable mass profile. Consequently, we adopt a Gaussian prior on the logarithmic slope, $\gamma_l = 2.08 \pm 0.13 $, measured by Shajib et al. (2021) \citep{Shajib2020} on the SLACS galaxies. We use a S\'ersic light-profile superimposed to a set of shapelets \citep{Refregier2003,Birrer2015} for the source galaxy and fix $n_{max} = 8$ as the maximum order of the shapelet basis. This ensures a good source reconstruction of the small-scale features at a fairly low computation cost. In our baseline model, we place a Singular Isothermal Ellipsoid (SIE) at the position of the quasar. This is required to fit the gravitational arcs down to the noise level and improves the BIC by 234 as compared with a model without this substructure (see Table \ref{tab:BICvalue}). The corresponding Einstein radius is directly inferred within a Bayesian framework and later converted into an estimate of the mass within it via the lensing relation : 
\begin{equation}
    \theta_E = \sqrt{\frac{4GM(\theta<\theta_E)}{c^2} \frac{D_{ls}}{D_{l}D_{s}}}, 
\end{equation}
where $D_{l}$ is the angular diameter distance to the lens, $D_{s}$ is the distance to the source and $D_{ls}$ is the distance between the lens and the source.

\subsection*{Pixelated source reconstruction}
\begin{figure}[h!]
    \centering
    \includegraphics[width=\textwidth]{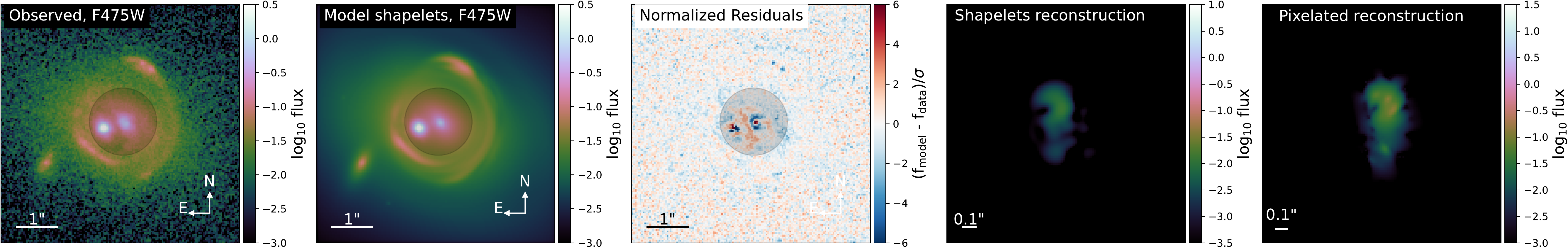}   
    \includegraphics[width=\textwidth]{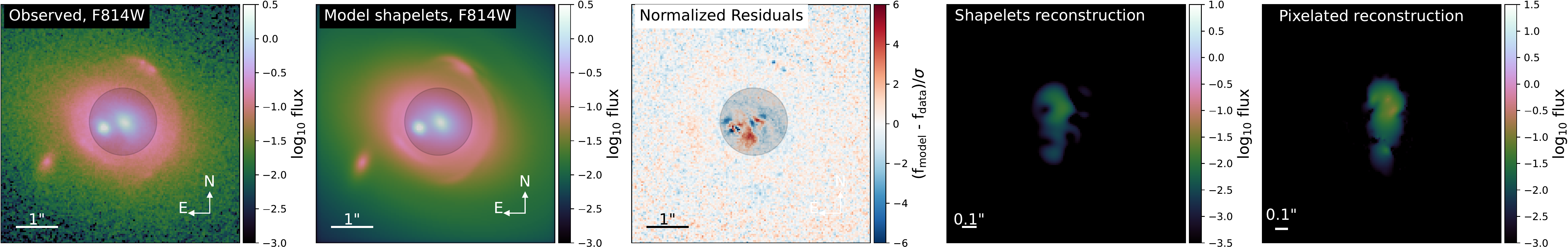}
    \caption{Lens model and source reconstruction of \Jzero in the HST F475W (top row) and F814W (bottom row) bands. \textit{From left to right:} HST images, best-fit lens model using shapelets source reconstruction, residuals of the fit, reconstructed source using shapelets and pixelated source reconstruction with \slit. The inner blue circle shows the mask used during the fit as the full complexity of the lens light distribution cannot be recovered with analytical profiles down to the noise level. These central pixels are still used in the alternative model to extract the host's photometry.}
    \label{fig:model}
\end{figure}

As an alternative to the reconstruction with shapelets and to further refine the source model, we performed a pixelated reconstruction of the source with the Python package \slit\ \cite{Joseph2019, galan2021}, from our best lens model. The source pixel values are regularized using priors based on sparsity, wavelet transforms and positivity constraint. The optimization of the cost function uses automatic differentiation. We use two types of wavelets including starlets, which are a set of basis functions that have been shown to be particularly suitable to represent the light profile of astronomical objects \cite{Starck2007}. The multi-resolution properties of the wavelets allow us to capture small scale features in the light profile that can not be fully modeled with shapelets. In addition, it ensures that the reconstructed source image contains only pixel values that are 3$\sigma$ above the noise level \citep[see][for details]{galan2021}. This technique allows us to provide high-resolution de-lensed images of the source galaxy located at redshift $z_s = 0.558$. The reconstruction is done separately in each band and we use the combination of the available bands to produce the color image of the source galaxy shown on the bottom right panel of Figure \ref{fig:rgb}. We also show in Figure \ref{fig:model} our best lens model, as well our shapelets and pixelated source reconstruction for each individual HST bands. 

\subsection*{Black hole mass estimation}  
Spectroscopic observations of \Jzero are available from the Sloan Digital Sky Survey (SDSS) seventh data release \cite{Abazajian2009} and from Keck/LRIS \cite{Courbin2012}. The broad H$\alpha$, H$\beta$ and \textsc{[Oiii]} emission lines of the quasar are visible in these data, making the redshift measurement of the quasar possible ($z_{l} = 0.209$). Also detected are the \textsc{[Oiii]} and \textsc{[Oii]} emission lines of the background start forming galaxy at  $z_s = 0.558$, both in the SDSS and in Keck/LRIS spectra. 

\begin{figure}[h!]
    \centering
    \includegraphics[width=0.5\textwidth]{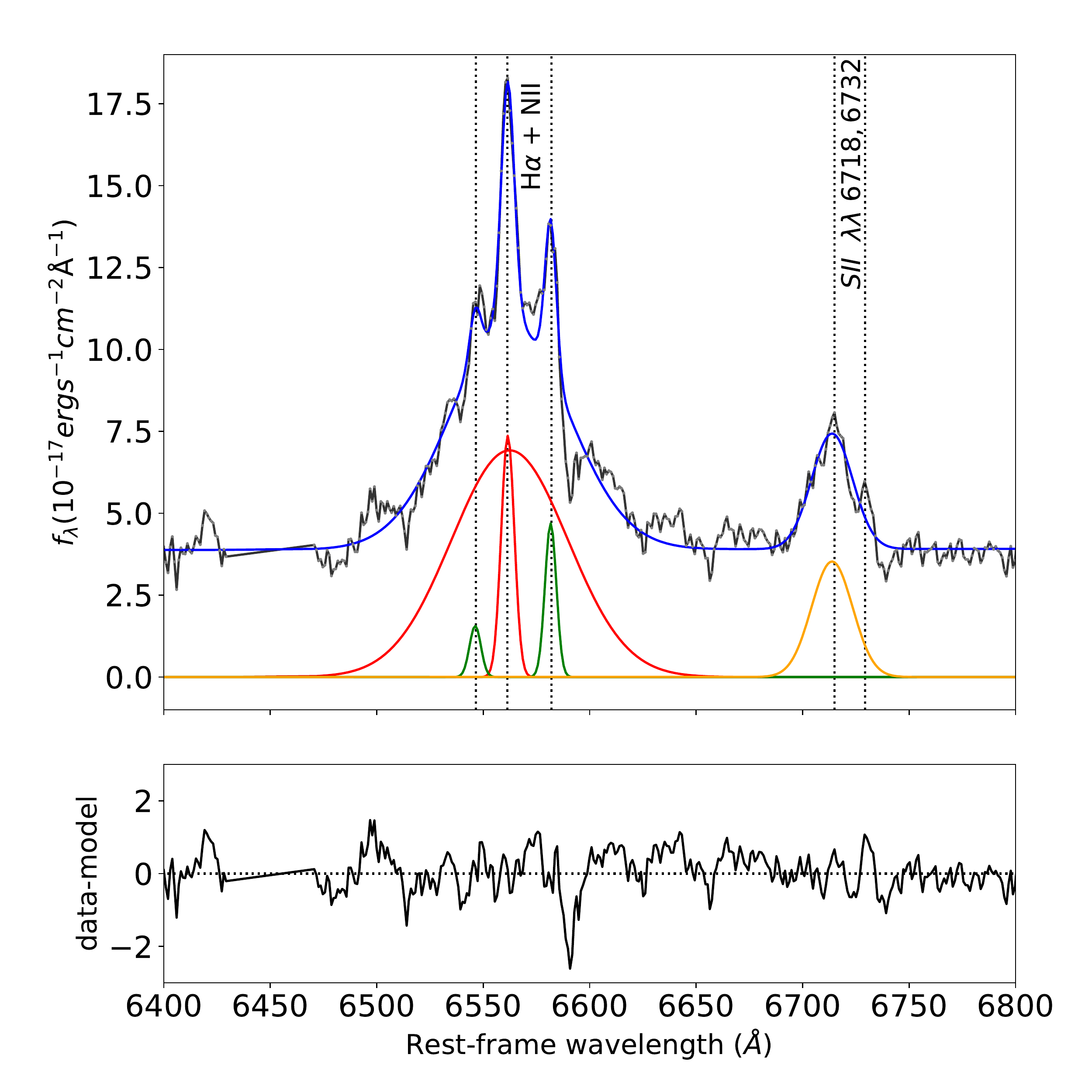}
    \caption{Local fit (blue) of the Keck spectrum (black) around the broad $H\alpha$ emission line. We include four components in our model: the continuum (yellow), a broad and a narrow component for the $H\alpha$ emission (red) and the \textsc{[Nii]} doublet (green). The bottom panel shows the residuals of the fit. }
    \label{fig:halpha_fit}
\end{figure}

To obtain an estimate of the black hole mass, we first subtract the host component from the SDSS and Keck spectra with {\tt pyQSOfit} \citep{Guo2018}, which enables a modeling of the host using a Principal Component Analysis (PCA) based on the host components from Yip et al. (2004)\citep{Yip2004}. In practice, rest-frame data below 3450 \angstrom are ignored when the host decomposition is performed as the basis eigenvectors from the PCA are not available at bluer wavelengths. We also correct for galactic extinction based on the extinction maps of Schlegel et al. (1998) \cite{Schlegel1998}. This is generally a small correction, i.e. amounting less than 20\% of the flux at the wavelength of H$\alpha$, but it matters for getting a more reliable host-QSO decomposition. Once a reliable host-QSO decomposition is achieved, we perform a local fit of the H$\alpha$ (which has better SNR than H$\beta$) on the host-corrected spectra. The result of this fit is shown in Figure \ref{fig:halpha_fit}. We considered the following two data sets: (1) the SDSS spectrum, which is relatively noisy but properly calibrated in flux. The fiber is 3" in diameter and therefore contains a large fraction of the host. (2) The Keck long-slit spectrum published in Courbin et al. 2011 \citep{Courbin2012}, which contains a smaller fraction of the host but is only corrected for instrumental response curve. Therefore, we obtain the black hole mass from the FWHM of the broad H$\alpha$ line measured on both the Keck and SDSS spectra and from the quasar luminosity derived only from SDSS. For the local fit of H$\alpha$, we perform a local fit of the continuum, and fit narrow emission lines for H$\alpha$, and for the \textsc{[Nii]} doublet. The separations between those 3 lines are forced to match the theoretical one, while their width is assumed to be identical. An additional broad component is assumed for H$\alpha$, but its zero velocity is not forced to be at the systemic velocity. A small offset by 40 km.s$^{-1}$ w.r.t. the systemic redshift is effectively measured. The scaling relation used to derive the black hole mass is the one proposed for a local fitting procedure presented in Mejia-Restrepo et al. (2015) \citep{Mejia2016}:
\begin{equation}
    M_{BH} = K(\lambda L_{\lambda})^{\alpha}(\mathrm{FWHM})^2 
\end{equation}
with L in units of 10$^{44}$ erg.s$^{-1}$ and FWHM in units of 1000 km.s$^{-1}$.  For H$\alpha$ line, we have $\log(K) = 6.779$ and $\alpha=0.634$. We use the luminosity at 6200\,\angstrom  derived from the de-redened and host-corrected SDSS spectrum.

We find a black hole mass of $\log (M_{BH}/M_{\odot}) = 7.29$ from the SDSS spectrum and $\log (M_{BH}/M_{\odot}) = 7.32$ from the Keck/LRIS data (see Table \ref{tab:M_BH}). We adopt a conservative 0.35 dex error, corresponding to the intrinsic scatter of the relation, which dominates over the other sources of error. These estimates are in good agreement with the estimates from Liu et al. (2019) \citep{Liu2019}, who derived $\log (M_{BH}/M_{\odot}) = 7.15$ from the $H\beta$ line and $\log (M_{BH}/M_{\odot}) = 7.20$ from H$\alpha$ using the relation described in Ho \& Kim. (2015) \cite{Ho2015}.

\begin{table}[h!]
    \centering
    \caption{\label{tab:M_BH} FWHM of the broad H$\alpha$ line, luminosity at 6200\angstrom and black hole mass estimates of \Jzero.} 
\begin{tabular}{l|l|c|c|c}
Object     & Spectrum                  & \multicolumn{1}{l|}{FWHM (km/s)} & \multicolumn{1}{l|}{$L_{6200}$ {[}$10^{44}$ erg/s{]}} & \multicolumn{1}{l}{$\log (M_{BH}/M_{\odot})$} \\ \hline
\Jzero & SDSS                      & 2840                             & 0.24                                                & 7.29                                         \\
\Jzero & Keck                      & 2940                             & 0.24                                                & 7.32                                         \\
\end{tabular}
\end{table}

We note that the FWHM derived for H$\alpha$ from the local fit depends little of the data quality and of the host correction. We find agreement within 100 km/s between SDSS and Keck spectra, the host spectrum being subtracted or not. We note here that while the PCA predicts emission/absorption at the level of H$\alpha$ line, only flat continuum emission is subtracted, such that our H$\alpha$ model effectively models narrow emission from the host and from the AGN's narrow line region.

\subsection*{Stellar-mass estimation and modeling of the stellar velocity dispersion}
Photometric measurements of the quasar host galaxy come as a by-product of our lens modeling procedure. To ensure that the best fit of the lens light components is reached, we remove the central mask from our baseline model and rerun the optimization with all degrees of freedom in the lens light models, in addition to all the mass and source parameters. We optimize a total of 103 linear parameters and 62 non-linear parameters. When all pixels are included in the calculation of the likelihood, we observe a marginal change of the posterior distribution of the lens mass parameters and we find $\log_{10}(M_{\rm Tot, h}/ M_{\odot}) = 10.23^{+0.03}_{-0.03}$, well compatible with the estimate from our baseline model. 
 
From this model, we extract the photometry of the host within an aperture of $0\farcs71$ in diameter using the double S\'ersic model fitted to the HST images. In doing so, contamination by the main lens galaxy, by the lensed source and by the quasar are removed. Using the HST zero-point for photometric calibration, we obtain a magnitude of 20.33 and 21.41 in the F814W and F475W bands respectively.  For comparison, the magnitude of the quasar, modeled as a point source, is 20.88 and 21.89 in the F814W and F475W bands respectively. The luminosity ratio between the quasar and its host galaxy is therefore around 0.6 in both bands. We adopt a conservative 0.2 mag photometric error on these estimates before proceeding to a fit of the spectral energy distribution (SED) of the quasar host galaxy. This reflects the worst measured discrepancy between the full lens modeling photometry and the aperture photometry of the host performed on the PSF-subtracted HST data.

The stellar mass of the quasar host galaxy, $M_{\star, h}$, is inferred with the SED fitting package GSF(\url{https://github.com/mtakahiro/gsf}) \citep{Morishita2019}, adopting a range of stellar ages up to 10.0 Gyr and solar metallicity. A single color measurement is not really sufficient to constrain the stellar template, the Star Formation Rate, the age, and the metallicity but the quantity we are interested in, i.e. the stellar mass, is mostly constrained by the normalization of the stellar template, which is well-measured in our two filters.  Our best fits to the measured flux in each of the two HST bands are obtained for stellar population aged of 5.2, 4.7 and 4.5 Gyr, i.e. $\log_{10}(M_{\star, h}/ M_{\odot}) = 10.10~$, $\log_{10}(M_{\star, h}/ M_{\odot}) = 9.80~$, and $\log_{10}(M_{\star, h}/ M_{\odot}) = 9.81~$ for Salpeter, Chabrier and Kroupa IMF respectively. The uncertainty of $M_{\star, h}$ is estimated to 0.2 dex \citep{Treu2007, Park2015, Ding2020} and includes the uncertainties due to the age/metallicity degeneracy \citep{Bell2001}. Changing the metalicity from solar metalicity $Z_{\odot}$ to 0.5 $Z_{\odot}$ or 1.6 $Z_{\odot}$ changes the central value of our mass estimates by less than 0.1 dex.

We repeat this procedure to estimate the stellar mass of the main galaxy, $M_{\star, m}$. This time, we remove the light components originating from the quasar and the quasar host galaxy before measuring the flux within an aperture of $2\farcs03$, corresponding to twice the Einstein radius of the main lens. We obtain a magnitude of 18.50 and 20.62 in the F814W and F475W bands. This translates to a stellar mass of $\log_{10}(M_{\star, m}/ M_{\odot}) = 10.84 ~$, $\log_{10}(M_{\star, m}/ M_{\odot}) = 10.72 ~$, and $\log_{10}(M_{\star, m}/ M_{\odot}) = 10.75 ~$, and stellar ages of 7.8, 9.5 and 10.0 Gyr for Salpeter, Chabrier and Kroupa IMF respectively. 

Finally, the lens model of \Jzero allows us to estimate the stellar velocity dispersion of both the main and quasar host galaxy. Here, we assume that the motion of the stars is isotropic within the Einstein radii of each mass component. From spherical Jeans modeling, we obtain a luminosity (F814W) weighted line-of-sight velocity dispersion \citep[see e.g.][]{Binney1987, Suyu2010} for the quasar host galaxy of $\sigma_{\star ,h} = 111 \pm 3$ \ks, within an aperture of $0\farcs71$, corresponding to the Einstein radius. For the main galaxy, we estimate the luminosity weighted line-of-sight velocity dispersion to be $\sigma_{\star ,m} = 227 \pm 3$ \ks, within an aperture of $2\farcs03$ and in the F814W filter. Note that these estimates are only indicative as they reflect the conversion of the lensing total mass in terms of dynamics.

\subsection*{Lens modeling robustness checks}

\paragraph{Mass model assumptions :} We verify that the mass estimate of the quasar host galaxy is robust against the assumptions made on the mass profile of the main lens galaxy and on the quasar host. From our baseline model, composed of a PEMD for the main galaxy and of an SIE for the quasar host, we tested if the fit is improved by relaxing the slope of the mass profile of the quasar host. The imaging reduced $\chi _{img}^2$ is indeed slightly improved by $\Delta \chi _{img}^2 = -1.08 \cdot 10^{-4}$, but the introduction of an extra degree of freedom degrades the BIC ($\Delta$BIC=4). The estimated mass within the Einstein radius of the host galaxy for these two models still remains compatible within $\sim 0.5 \sigma$. We perform similar tests for different mass modeling assumptions of the main galaxy (SIE or PEMD) and for the quasar host (SIE, PEMD and Point Mass). Table \ref{tab:BICvalue} summarizes the inferred host mass, the BIC, and the imaging $\chi_{img}^2$ of each of these modeling choices. Using the BIC as a proxy for Bayesian evidence, we can combine the 6 models of Table \ref{tab:BICvalue} which include the quasar host galaxy, weighted by their BIC. This gives $\log_{10}(M_{\rm Tot, h}/M_{\odot}) = 10.27^{+0.07}_{-0.07}~$, which is very similar to our baseline estimate. The weights attributed to each model are 1.00 for the first model (PEMD+SIE), 0.135 for the second model (PEMD+PEMD), 0.011 for the third model (SIE+SIE), and 0 for the other three models.

In our baseline model, perturbations of the lensing potential introduced by any object not explicitly modeled are captured by external shear. The only significantly bright object not explicitly in our model is the companion galaxy at the South-East of the main lens galaxy. If we explicitly include it in our models, assuming it is at the same redshift as the quasar, we find its mass being $\log_{10}(M_{\rm comp}/ M_{\odot}) < 8.41 ~$ at 99\% confidence level (CL). The mass estimate of the host galaxy remains compatible to our baseline model within 1$\sigma$. Additionally, we test the possibility of adding a point mass at the position of the quasar on top of a non-singular isothermal ellipsoid (NIE) to model the central black hole atop a cored bulge component. The data constrain the core size of the NIE to $r_{\rm core} < 0.30$ kpc at 99\%CL. We also obtain an upper limit for the point mass model representing the central black hole $\log_{10}(M_{BH}/ M_{\odot}) < 9.03 $ at 99\%CL.

\begin{table}[h!]
\centering
\renewcommand{\arraystretch}{1.3}
\caption{\label{tab:BICvalue} $\Delta BIC$, total mass of the quasar host galaxy, $M_{\rm Tot, h}$, and $\Delta \chi^{2}_{img}$ for different lens modeling assumptions. $\Delta BIC$ and $\Delta \chi^{2}_{img}$ are computed relatively to our baseline model that describes the main lensing galaxy as a PEMD and the quasar host as a SIE.}
\begin{tabular}{l|l|l|c|c|c}
\multicolumn{1}{c|}{\begin{tabular}[c]{@{}c@{}}Model main \\ lens galaxy\end{tabular}} &
  \multicolumn{1}{c|}{\begin{tabular}[c]{@{}c@{}}Model quasar\\  host galaxy\end{tabular}} &
  \multicolumn{1}{c|}{\begin{tabular}[c]{@{}c@{}}Model companion \\ galaxy\end{tabular}} &
  $\Delta$BIC &
  $\log_{10} \left(\frac{M_{\rm Tot, h}}{M_{\odot}}\right)$ &
  $\Delta   \chi^{2}_{img}$ \\ \hline
PEMD & SIE                    & -    & 0   & $10.27^{+0.06}_{-0.07}$ & 0.00 \\
PEMD & PEMD                   & -    & 4   & $10.30^{+0.07}_{-0.05}$  & -1.08 $\cdot 10^{-4}$ \\
SIE  & SIE                    & -    & 9   & $10.03^{+0.05}_{-0.06}$  & 5.72$\cdot 10^{-4}$  \\
PEMD & NIE + Point Mass & -   & 23  & $10.27^{+0.07}_{-0.08}$  & 5.07$\cdot 10^{-5}$  \\
PEMD & Point Mass                     & -    & 25  & $10.08^{+0.06}_{-0.06}$ & 7.07$\cdot 10^{-4}$  \\
PEMD & SIE                    & SIE  & 66  & $10.17^{+0.15}_{-0.16}$ & -1.80$\cdot 10^{-4}$ \\
PEMD & -                      & -    & 234 & -                       & 6.35$\cdot 10^{-3}$
\end{tabular}
\end{table}

\begin{figure}
    \centering
    \includegraphics[width=0.9\textwidth]{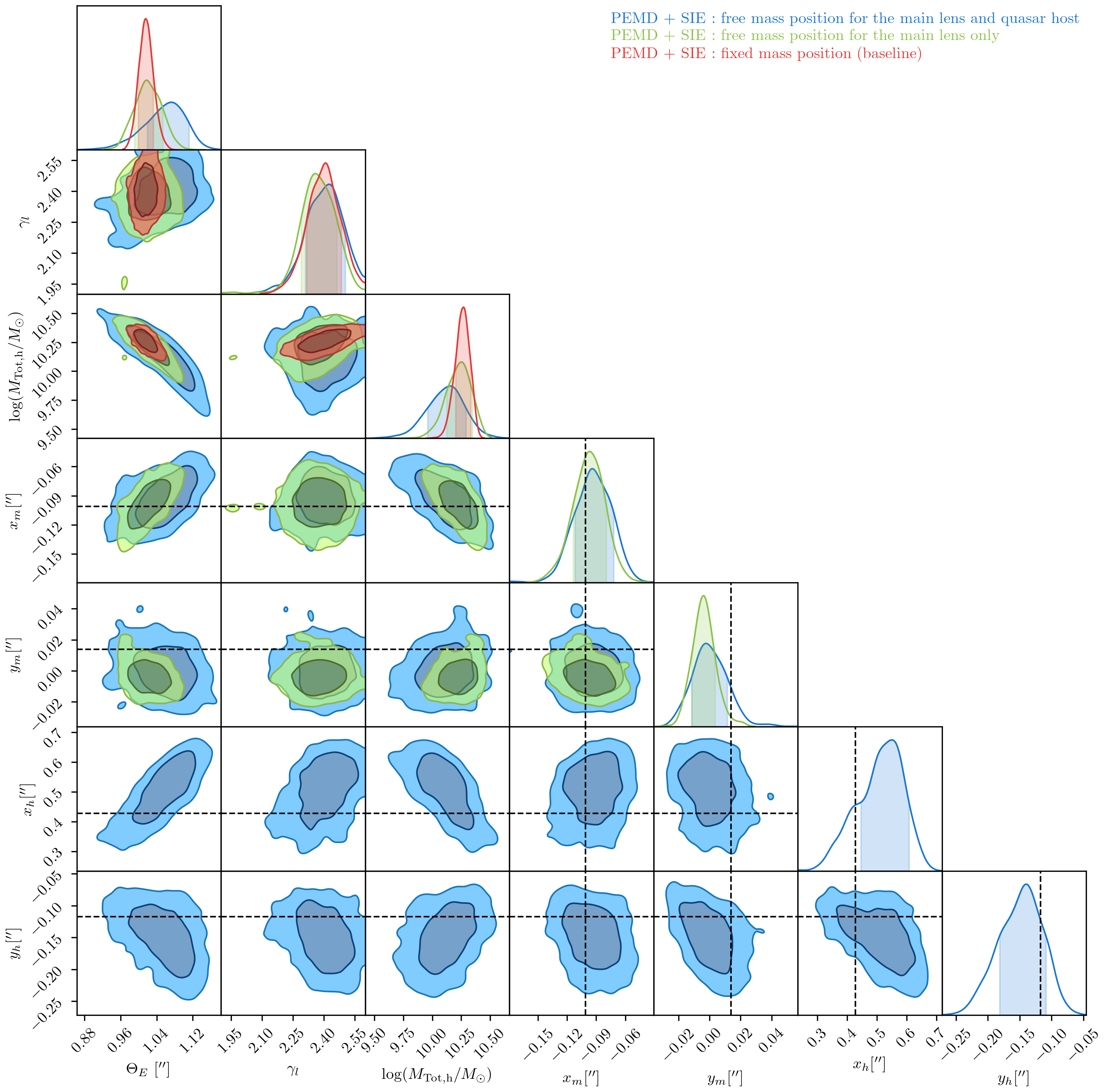}
    \caption{Posterior distributions (1$\sigma$ and 2$\sigma$ error contours) of a subset of the lens model parameters: the Einstein radius of the main galaxy $\Theta_E$, the slope of the PEMD model for the main lens, $\gamma_l$, the mass within Einstein radius of the quasar host galaxy $M_{\rm Tot, h}$, the mass centroid position of the main lens ($x_{m}$, $y_{m}$) and of the quasar host ($x_{h}$, $y_{h}$), in arcseconds. Our baseline model with mass profile tied to the light profiles is shown in red. The green and blue contours correspond to models where the mass centroid position is left free for the main lens and for both the main lens and the quasar host. The dashed lines indicate the astrometric positions of the center of light profiles. }
    \label{fig:mass_centroid}
\end{figure}

\paragraph{Light traces mass assumption :} We verify here that the light and mass profiles share the same centroids. To do this, we relax the assumption that light traces mass in our baseline model and we optimize all model parameters assuming independent centroids for the mass and light profiles of both the main lens and the quasar host galaxy. The posterior distribution of a subset of the modeling parameters is shown in Figure \ref{fig:mass_centroid}. 

When the centers of the mass profiles are allowed to vary during the fit, they naturally tend to align with the position of the light profiles within uncertainties. The inferred mass of the host is, with no surprise, less precise by 0.07dex; $\log_{10}(M_{\rm Tot, h}/ M_{\odot}) = 10.17 \pm 0.14$, but remains compatible with our baseline model within 1$\sigma$ uncertainties. This leaves all our conclusions unchanged and indicates that the light and mass distributions of the different components in our models share the same centroid.

\newpage
\bibliography{sample}
\section*{Data availability}
The HST images supporting this work are publicly available on the Hubble Legacy Archive (\url{https://hla.stsci.edu/}). Our reduced Keck and SDSS spectra are available on Zenodo (\url{10.5281/zenodo.7806468}). 

\section*{Code availability}
The lens modeling code \lenstro \ and the source reconstruction software \slit \ are freely accessible at \url{https://github.com/sibirrer/lenstronomy} and \url{https://github.com/aymgal/SLITronomy}. Stellar masses were estimated from the public python package {\tt GSF} (\url{https://github.com/mtakahiro/gsf}). The HST PSF was reconstructed using {\tt AstroObjectAnalyser} publicly available at \url{https://github.com/sibirrer/AstroObjectAnalyser}. Spectra have been fitted using {\tt pyQSOfit}, also publicly available at \url{https://github.com/legolason/PyQSOFit}.

\section*{Acknowledgements}
 M.M. acknowledges the support of the Swiss National Science Foundation (SNSF) under grant P500PT\_203114. M.M., F.C. and A.G. are supported by the European Research Council (ERC) under the European Union’s Horizon 2020 research and innovation programme (COSMICLENS: grant agreement No 787886) and the Swiss National Science Foundation (SNSF) under grant 200020\_200463. X.D. is supported by JSPS KAKENHI grant No. JP22K14071.

\section*{Author contributions statement}
M.M. conducted the analysis. A.G. developed the source reconstruction algorithm \slit. M.M. and F.C wrote the manuscript, X.D. did the stellar population analysis. D.S. measured the black hole mass. All other co-authors have actively participated in the discussions, in the HST data acquisition and in the discovery process of \Jzero. 

\section*{Competing Interests Statement}

The authors declare no competing interest.

\end{document}